\documentclass[12pt,showpacs,preprintnumbers,superscriptaddress,amsmath,amssymb,nofootinbib]{revtex4}
\usepackage{graphicx}
\usepackage{dcolumn}
\usepackage{bm}
\usepackage{amssymb}
\usepackage{amsmath}
\usepackage{epsfig}    
\usepackage{color}
\usepackage{hhline}

\def\be{\begin{equation}}
\def\ee{\end{equation}}
\newcommand{\bea}{\begin{eqnarray}}
\newcommand{\eea}{\end{eqnarray}}
\newcommand{\nn}{\nonumber}

\numberwithin{equation}{section}

\begin{document}

\title{Classically Conformal Radiative Neutrino Model \\ with\\ Gauged $B-L$ Symmetry}
\preprint{KIAS-P14078}
\author{Hiroshi Okada}
\email{hokada@kias.re.kr}
\affiliation{School of Physics, KIAS, Seoul 130-722, Korea}

\author{Yuta Orikasa}
\email{orikasa@kias.re.kr}
\affiliation{School of Physics, KIAS, Seoul 130-722, Korea}
\affiliation{Department of Physics and Astronomy, Seoul National University, Seoul 151-742, Korea}

\date{\today}

\begin{abstract}
We propose a classically conformal model in a minimal radiative seesaw, in which we employ a gauged $B-L$ symmetry 
in the standard model
that is essential in order to  work the Coleman-Weinberg mechanism well that induces the $B-L$ symmetry breaking. 
As a result, 
nonzero Majorana mass term and
electroweak symmetry breaking simultaneously occur. In this framework,
we show a benchmark point 
to satisfy several theoretical and experimental constraints.
Here theoretical constraints represent inert conditions and Coleman-Weinberg condition.
Experimental bounds come from lepton flavor violations (especially  $\mu\to e\gamma$), the current bound on the $Z'$ mass at the CERN Large Hadron Collider, and neutrino oscillations.
\end{abstract}
\maketitle
\newpage

\section{Introduction}
 Nowadays the standard model (SM) becomes trustworthy to describe microscopic fundamental physics, since the SM Higgs has been discovered at the CERN Large Hadron Collider (LHC).
 However it has to still be extended in order to include dark matter (DM) candidate and
 tiny but massive neutrinos whose existences are indirectly or directly shown by several experimental evidences.
{Radiative seesaw models are one of the sophisticated solutions to explain both issues simultaneously, where new fields have to be introduced as mediators in the loop of the neutrino masses.}
{One of such exotic fields can frequently be identified as a DM candidate, when it is neutral under the electric charge. Then} neutrinos have a correlation to the DM candidate.
Due to the fascinating nature, there  exists a vast number of  papers along this idea~\cite{Zee,Li, Zee-Babu, Krauss:2002px, Ma:2006km,
Hambye:2006zn,
Aoki:2013gzs, Dasgupta:2013cwa, Aoki:2008av, MarchRussell:2009aq,
Schmidt:2012yg, Bouchand:2012dx, Aoki:2011he, Farzan:2012sa, 
Bonnet:2012kz, Kumericki:2012bf, Kumericki:2012bh, Ma:2012if, Gil:2012ya,
Okada:2012np, Hehn:2012kz, Dev:2012sg, Kajiyama:2012xg, Okada:2012sp,
Aoki:2010ib, Kanemura:2011vm, Lindner:2011it, Kanemura:2011mw,
Kanemura:2012rj, Gu:2007ug, Gu:2008zf, Gustafsson, Kajiyama:2013zla, Kajiyama:2013rla,
Hernandez:2013dta, Hernandez:2013hea, McDonald:2013hsa, 
Baek:2013fsa, Ma:2014cfa,  Ahriche:2014xra,
Kanemura:2011jj, Kanemura:2013qva, Kanemura:2014rpa,
Chen:2014ska, Ahriche:2014oda, Okada:2014vla, Ahriche:2014cda,
Aoki:2014cja, Lindner:2014oea,Davoudiasl:2014pya, Ahn:2012cg, Ma:2012ez,
Kajiyama:2013lja, Kajiyama:2013sza, Ma:2013mga, Ma:2014eka, radlepton1,radlepton2,radlepton3, Schmidt:2014zoa, Long1,Okada:2014oda, Long2, Nebot,
Fraser:2014yha, Okada:2014qsa, Ma:2014yka, Sierra:2014rxa,Ma:2014qra}. 
Especially, Ma model~\cite{Ma:2006km} is renowned as  {one of the} minimal radiative seesaw models including fermionic or bosonic DM candidate.

 As another aspect to be resolved in the SM context, there exists the hierarchy problem.
One of the {popular solutions} 
is to extend to be supersymmetrized,
but one cannot hitherto find  any signals {at LHC. 
Thus several alternative solutions {have been} discussed~\cite{Foot:2007as, Foot:2007ay, 
Foot:2007iy, Shaposhnikov:2008xi, Aoki:2012xs, Hamada:2012bp, Farina:2013mla, Heikinheimo:2013fta, 
Giudice:2013yca, Tavares:2013dga, Kawamura:2013kua, Haba:2013lga, Abel:2013mya,Dorsch:2014qpa,  Ibe:2013rpa, 
Kobakhidze:2014afa, Bardeen:1995kv} in these days.
Here we focus on a new approach inspired by Bardeen's argument~\cite{Bardeen:1995kv}}.
He suggests that once the classically conformal symmetry and its minimal violation 
by quantum anomalies are imposed on SM, it may be free from quadratic divergences. 
 Such theories based on this idea are known as classically conformal models~\cite{Hempfling:1996ht, Meissner:2006zh, 
Espinosa:2007qk, Espinosa:2008kw, Chang:2007ki, Iso:2009ss, Iso:2009nw, Holthausen:2009uc, AlexanderNunneley:2010nw, 
Hur:2011sv, Ishiwata:2011aa, Iso:2012jn, Oda:2013rx, Englert:2013gz, Hambye:2013dgv, Khoze:2013oga, Carone:2013wla, 
Farzinnia:2013pga, Gabrielli:2013hma, Hashimoto:2013hta, Guo:2014bha, Hashimoto:2014ela, Radovcic:2014rea,
Khoze:2014xha, Tamarit:2014dua, Lindner:2014oea, Altmannshofer:2014vra, Benic:2014aga, Kang:2014cia},
in which  any mass terms are forbidden but all dimensional parameters( including mass terms) are dynamically generated in the classical Lagrangian.
Due to absence of any intermediate scales between the TeV scale and Planck scale,
the Planck scale physics can directly be connected to the electroweak (EW) physics.  
 Once we combine the classically conformal model with a radiative seesaw(such as Ma model),
the model potentially has a direct connection between tiny neutrino mass scale(eV) and Planck scale due to the conformal nature.
However Ma model with the classically conformal symmetry cannot be realistic because of the  following two reasons.
The first one is that the EW symmetry breaking doesn't occur due to the largeness of top Yukawa coupling.
The second one is that the classically conformal symmetry forbids Majorana mass term that plays an important role in  generating neutrino masses.
In order to resolve these two problems, we employ a gauged $B-L$ model as a minimal extension of Ma model  in this paper.
Then the EW symmetry breaking is triggered by $B-L$ symmetry breaking and 
Majorana mass term is arisen by the $B-L$ symmetry breaking.

This paper is organized as follows.
In Sec.~II, we show our model building including neutrino mass.
In Sec.~III, we show our numerical results. We conclude in Sec.~VI.

\section{The Model}


\begin{table}[thbp]
\centering {\fontsize{10}{12}
\begin{tabular}{|c||c|c|c|}
\hline Fermion & $L_L$ & $ e_{R} $ & $N_R$  
  \\\hhline{|=#=|=|=|$}
$(SU(2)_L,U(1)_Y)$ & $(\bm{2},-1/2)$ & $(\bm{1},-1)$ & $(\bm{1},0)$  
\\\hline
$U(1)_{B-L} $ & $-1$ & $-1$ &  $-1$    \\\hline
$Z_2$ & $+$ & $+$ &  $-$  \\\hline
\end{tabular}%
} \caption{Fermion sector; notice the three flavor index of each field $L_L$, $e_R$, and $N_R$ is abbreviated.} 
\label{tab:1}
\end{table}

\begin{table}[thbp]
\centering {\fontsize{10}{12}
\begin{tabular}{|c||c|c|c|}
\hline Boson  & $\Phi$   & $\eta$    & $\varphi$ 
  \\\hhline{|=#=|=|=|}
$(SU(2)_L,U(1)_Y)$ & $(\bm{2},1/2)$  & $(\bm{2},1/2)$   & $(\bm{1},0)$ \\\hline
$U(1)_{B-L} $ & $0$ & $0$ &  $2$    \\\hline
$Z_2$ & $+$ & $-$ &  $+$  \\\hline
\end{tabular}%
} 
\caption{Boson sector }
\label{tab:2}
\end{table}

In this section, we devote to review our model, where
the particle contents for fermions and bosons are respectively shown in Tab.~\ref{tab:1} and Tab.~\ref{tab:2}. 
We add three Majorana fermions $N_R$ with isospin singlet  but $-1$ charge under the gauged $B-L$ symmetry to the SM fields.
Notice here the number of 'three' flavors to $N_R$ is uniquely determined by the anomaly cancellation of the gauged $B-L$ symmetry.
 For new bosons, we introduce
 a neutral isospin singlet scalar $\varphi$ with  $+2$ charge under the $B-L$ symmetry. 
The other bosons $\eta$ and $\Phi$ are neutral under the $B-L$ charge.
Then we assume that  the SM-like Higgs $\Phi$ and the gauge single $\varphi$ have  vacuum
expectation value (VEV); $v/\sqrt2$ and $v'/\sqrt2$, respectively,
where the VEV of $\varphi$ spontaneously breaks  the $B-L$ symmetry down. 
Even after the breaking of $B-L$ symmetry as well as electroweak symmetry, a remnant discrete  symmetry $Z_2$ remains.
This $Z_2$ symmetry plays a role in  assuring the stability of our  DM candidate.

The relevant  Lagrangian for Yukawa sector and scalar potential under these assignments
are given by
\begin{eqnarray}
-\mathcal{L}_{Y}
&=&
(y_\ell)_a \bar L_{La} \Phi e_{Ra} + (y_{\eta})_{ab} \bar L_{La} \eta^*   N_{Rab}
+\frac12 (y_{N})_a \varphi \bar N^c_{Ra} N_{Ra} 
+\rm{h.c.} \label{Lag:Yukawa}\\ 
\mathcal{V}
&=& 
  \lambda_\Phi |\Phi|^4 
  + \lambda_\eta |\eta|^4 
  + \lambda_\varphi |\varphi|^4
  + \lambda_{\Phi\eta} |\Phi|^2 |\eta|^2
  + \lambda'_{\Phi\eta}  |\Phi^\dagger \eta|^2
  + \lambda''_{\Phi\eta}  [(\Phi^\dag\eta)^2+{\rm c.c.}]\nn\\
  &&
  + \lambda_{\Phi\varphi}  |\Phi|^2 |\varphi|^2 + \lambda_{\eta\varphi} | \eta|^2 |\varphi|^2
,
\label{HP}
\end{eqnarray}
where each of the index $a$ and $b$ that runs $1$ to $3$ represents the number of generations,  and the first term of $\mathcal{L}_{Y}$ generates the diagonal charged-lepton mass matrix. Notice here that any mass terms are forbidden by the conformal symmetry. 
Without loss of generality, we can work on the basis where $y_{N}$ is diagonal matrix with real and positive.

\subsection{Symmetry breaking}
\label{sec:SB}
In this subsection, we explain how the symmetry breaking occurs in our model, where
the RGEs related to the breaking are given in the Appendix. 
First of all we impose the classically conformal symmetry to our model.
Then the EW symmetry breaking occurs not by negative mass parameter but by radiatively,
because of absence of any kind of mass terms.
%
Furthermore we assume the following conditions at the Planck scale as simple as possible in our theory, 
\begin{eqnarray}
\lambda_{\Phi\eta}=\lambda_{\Phi\eta}^\prime=\lambda_{\Phi\varphi}=\lambda_{\eta\varphi}=0. 
\label{eq:conditions}
\end{eqnarray}
In principle, all the quartic couplings except $\lambda_\varphi$ and $ \lambda''_{\Phi\eta}$ can be zero.
\footnote{Nonzero $\lambda_\varphi$ is minimally required in order to work Coleman-Weinberg mechanism 
in the $B-L$ model sufficiently~\cite{Hashimoto:2013hta, Hashimoto:2014ela}. 
When $ \lambda''_{\Phi\eta}$ is zero at Planck scale, the coupling has to be zero at all the scale as can be seen in Eq.(A.12). It suggests that the neutrino masses are zero, which is not experimentally allowed. 
}
However, we assume $\lambda_\Phi$ and $\lambda_\eta$ to be nonzero
for the following technical reasons: 
nonzero $\lambda_\Phi$ plays an important role in obtaining the SM Higgs mass, 
and nonzero $\lambda_\eta$ is required by the inert condition as you will see in the next subsection. 
Under these assumptions, these couplings in Eq.~(\ref{eq:conditions}) are generated by quantum correction. As a result,  these couplings 
are very small at low energy scale.  
Therefore we can consider the $B-L$ 
sector and the SM with inert doublet sector separately.

At first we consider the $B-L$ sector. 
The $B-L$ symmetry is broken by the Coleman-Weinberg mechanism~\cite{Coleman:1973jx}.
{And the running coupling 
$\lambda_{\varphi}$ (and related parameters $g_{B-L}, y_N$) should satisfy the following relation
at the $B-L$ symmetry breaking scale ($v'$),
\begin{align}
\lambda_{\varphi}(\mu=v') \sim \frac{3}{4\pi^2}\left(g^4_{B-L} 
 -\frac{1}{96}Tr\left[y_N^\dagger y_N y_N^\dagger y_N\right] \right).
 \label{CWcon}
\end{align}
}
{Thus}
the mass of  $\varphi$ is obtained by the following form, 
\begin{eqnarray}
 m_\varphi^2=-4\lambda_\varphi v'^2. 
\end{eqnarray}

Once the $B-L$ symmetry is broken, the mass of SM-like Higgs is induced through the mixing 
between the SM Higgs ($\Phi$) and $B-L$ breaking scalar ($\varphi$) in the potential. Therefore the effective tree-level mass squared is arisen.
Remind here that the EW symmetry breaking occurs in the same way as SM if $\lambda_{\Phi\varphi}$ is negative.
In our case, the negative $\lambda_{\Phi\varphi}$ arises from our RGE(see Eq. (\ref{RGEphivphi})) with positive sign under our assumption($\lambda_{\Phi\varphi}(M_{pl})=0$). 
Finally, inserting the tadpole condition; $\lambda_\Phi=-\lambda_{\Phi\varphi} v'^2/(2v^2)$, the mass of SM-like Higgs is given by 
\begin{align}
m_h^2 = -\lambda_{\Phi\varphi}(\mu=v')v'^2. 
\end{align}

\subsection{Scalar sector}
After the EW symmetry breaking, each of scalar field has nonzero mass.
We parametrize  these scalar fields as 
\begin{align}
&\Phi =\left[
\begin{array}{c}
\phi^+\\
\phi^0
\end{array}\right],\
\eta =\left[
\begin{array}{c}
\eta^+\\
\eta^0
\end{array}\right].
\label{component}
\end{align}
And the neutral components of the above fields and the singlet scalar field can be expressed as
\begin{align}
\phi^0=\frac1{\sqrt2}(v+h),\ 
\varphi = \frac1{\sqrt2}(v'+\rho),
\label{Eq:neutral}
\end{align}
where 
 $v$ is written in terms of 
the Fermi constant $G_F$ by $v^2=1/(\sqrt{2}G_F)\approx(246$ GeV)$^2$.

$\eta$ is the inert doublet and the mass of $\eta$ should be positive.  
In our model, the $\eta$ mass is generated {through the quartic term of $\lambda_{\eta\varphi}$.
Consequently, the term} should be positive at the symmetry breaking scale, 
\begin{eqnarray}
\lambda_{\eta\varphi}>0. 
\label{inert1}
\end{eqnarray}
In addition, the quartic couplings satisfy the following inert conditions~\cite{Barbieri:2006dq}, 
\begin{eqnarray}
\lambda_\Phi>0,\  \lambda_\eta>0,\  
\lambda_{\Phi\eta}+\lambda'_{\Phi\eta}-\mid\lambda''_{\Phi\eta}\mid>-2\sqrt{\lambda_\Phi\lambda_\eta}. 
\label{inert2}
\end{eqnarray}

The mass matrix of the neutral component of $h$ and 
$\rho$ is given by
\begin{equation}
m^{2} (h,\rho) = \left(%
\begin{array}{cc}
 -\lambda_{\Phi\varphi} v'^2 & \lambda_{\Phi\varphi} vv' \\
 \lambda_{\Phi\varphi}vv' & -4\lambda_{\varphi} v'^2 \\
\end{array}%
\right) \!=\! \left(\begin{array}{cc} \cos\theta & \sin\theta \\ -\sin\theta & \cos\theta \end{array}\right)
\left(\begin{array}{cc} m^2_{h_{\rm SM}} & 0 \\ 0 & m^2_{H}  \end{array}\right)
\left(\begin{array}{cc} \cos\theta & -\sin\theta \\ \sin\theta &
      \cos\theta \end{array}\right), 
\end{equation}
where $h_{\rm SM}$ is the SM Higgs and $H$ is an additional Higgs mass
eigenstate. The mixing angle $\theta$ is given by 
\be
\tan 2\theta=\frac{ -2\lambda_{\Phi\varphi}v v'}{v'^2(4\lambda_\varphi - \lambda_{\Phi\varphi})}.
\label{II.12}
\ee
{Therefore} $h$ and $\rho$ are rewritten in terms of the mass eigenstates $h_{\rm SM}$ and $H$ as
\begin{eqnarray}
h &=& h_{\rm SM}\cos\theta + H\sin\theta, \nn\\
\rho &=&- h_{\rm SM}\sin\theta + H\cos\theta.
\label{eq:mass_weak}
\end{eqnarray}
The mixing angle $\theta$ is generally constrained by $h_{\rm SM} \rightarrow \gamma\gamma$ process at LHC.
{But} we can avoid such a constraint, since  we expect $v/v'\le(0.1)<<1$ as can be seen in Eq.(\ref{II.12}).
The other scalar masses are found as 
\begin{eqnarray}
m_\eta^2\equiv m^{2} (\eta^{\pm}) &=&  \frac12 (\lambda_{\Phi\eta} v^{2} + \lambda_{\eta\varphi} v'^{2}),\\ 
m^2_{R}\equiv m^{2} (\mathrm{Re}\:\eta^{0}) &=&
\frac12 \left[(\lambda_{\Phi\eta}+\lambda'_{\Phi\eta} + 2\lambda''_{\Phi\eta}) v^{2} +  \lambda_{\eta\varphi} v'^{2}\right], \\ 
m^2_{I}\equiv m^{2} (\mathrm{Im}\:\eta^{0}) &=& 
\frac12 \left[(\lambda_{\Phi\eta}+\lambda'_{\Phi\eta} - 2\lambda''_{\Phi\eta}) v^{2} +  \lambda_{\eta\varphi} v'^{2}\right].
\end{eqnarray}
{Notice here that there exists a constraint between $m_\eta$
 and $m_{I}$~\footnote{We assume $m_{\eta_I}$ is lighter than $m_{\eta_R}$, {\it i.e.}, $\lambda''_{\Phi\eta}$ is positive.} that comes from the $S$-$T$-$U$ parameter~\cite{Barbieri:2006dq}.


\subsection{Neutrino mass matrix}
The neutrino mass matrix is obtained at one-loop level as follows~\cite{Ma:2006km, Hehn:2012kz}:
\bea
({\cal M}_\nu)_{ab}=
\frac{(y_{\eta})_{ak}(y_{\eta})_{bk}M_k}{(4\pi)^2}
\left[\frac{m^2_R}{m^2_R-M^2_k}\ln\frac{m^2_R}{M^2_k}
-\frac{m^2_I}{m^2_I-M^2_k}\ln\frac{m^2_I}{M^2_k}
\right],
\eea
where $M_k\equiv (y_N)_kv'/\sqrt2$ ($k=1-3$).
In this form, observed neutrino mass differences and their mixings 
are obtained through the Ref.~\cite{Hehn:2012kz} with a sophisticated way,
when {the charged-lepton mass matrix is diagonal}. 
Following this method,
{$y_\eta$} 
is generally written as
\begin{eqnarray}
{y_\eta} 
=U^*_{MNS}
\left(
\begin{array}{ccc}
m_1^{\frac{1}{2}} & 0 & 0 \\
0 & m_2^{\frac{1}{2}} & 0 \\
0 & 0 & m_3^{\frac{1}{2}} \\
\end{array} 
\right)
O R{^{-\frac{1}{2}}}, 
\label{yukawa}
\end{eqnarray}
where $U_{MNS}$ is the Maki-Nakagawa-Sakata (MNS) matrix, and $m_i$'s are neutrino mass eigenvalues.
$O$(that is an complex orthogonal matrix), and $R$(that is a diagonal matrix), are respectively formulated as
\begin{eqnarray}
O=
\left(
\begin{array}{ccc}
0 & 0 & 1 \\
\cos\alpha & \sin\alpha & 0 \\
-\sin\alpha & \cos\alpha & 0 \\
\end{array} 
\right),
\label{co}\quad \alpha\ {\rm is\ a \ complex\ parameter},
\end{eqnarray}
and
\begin{eqnarray}
R_{ii}=M_i\left(\frac{m_R^2}{m_R^2-M_i^2}\ln\frac{m_R^2}{M_i^2}-\frac{m_I^2}{m_I^2-M_i^2}\ln\frac{m_I^2}{M_i^2}\right).
\end{eqnarray}
Notice here that we assume the lightest neutrino mass is zero and the neutrino mass spectrum is normal hierarchy. 
In this case, one column of Yukawa matrix is zero.

\section{Numerical results}
In general aspect, VEV can be stable only when $\lambda_\varphi$ is negative as can be seen 
in Eq.(II.4) and Eq.(II.5).
We numerically solve the RGEs and find parameters that satisfy the inert conditions, 
Eq. (\ref{inert1}) and (\ref{inert2}).
Here we focus on calculating $\alpha=0$ (in Eq. (\ref{co})) case, because this case is one of the simplest way to satisfy the Lepton Flavor Violation (LFV) 
in Eq.~(\ref{yukawa}).~\footnote{
In general, the larger value of the imaginary part of $\alpha$ gives the larger Yukawa couplings. Therefore it becomes to be difficult to satisfy the LFV processes.
}
{The most stringent experimental upper bound comes from $\mu\to e\gamma$ process. 
Its branching ratio is calculated as
 \begin{eqnarray}
  {\rm Br}(\mu\to e\gamma)
&\!\!\!=\!\!\!&
\frac{3\alpha_{\mathrm{em}}}{64\pi(G_{F}m^{2}_{\eta})^{2}}
\left|\sum_{k=1}^3\left( y_{\eta}^{\dag}\right)_{\alpha{k}}
(y_{\eta} )_{k\beta}
F_2\left(\frac{M_k^2}{m_{\eta}^2}\right)\right|^{2},
 \label{LFV1}
 \end{eqnarray}
where $\alpha_{\mathrm{em}}=1/137$,
and the loop function $F_{2}(x)$ is given by
\begin{eqnarray}
F_{2}(x)=\frac{1-6x+3x^{2}+2x^{3}-6x^{2}\ln{x}}{6(1-x)^{4}}.
\end{eqnarray}
}

We use the following parameters at the Planck scale, 
\begin{eqnarray}
\lambda_\Phi=0.01, \  \lambda_\eta=0.09 ,\  \lambda_\varphi=0.011 ,\  \lambda_{\Phi\eta}''=10^{-9}, \ 
g_{B-L}=0.17 ,\  y_m=0.2. 
\end{eqnarray}
The RG flows of the quartic couplings are depicted in Fig.\ref{rge1}, Fig.\ref{rge2}, and Fig.\ref{rge3}. 
\begin{figure}[bt]
\begin{center}
\includegraphics[scale=0.7]{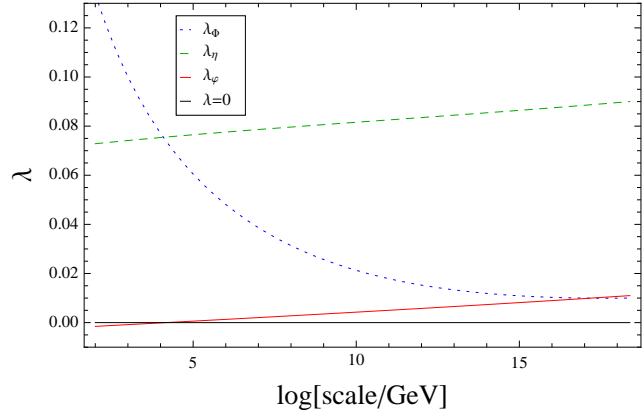}
   \caption{Running for quartic couplings. Black solid line is $\lambda=0$ axis. }
   \label{rge1}
\end{center}
\end{figure}
\begin{figure}[bt]
\begin{center}
\includegraphics[scale=0.7]{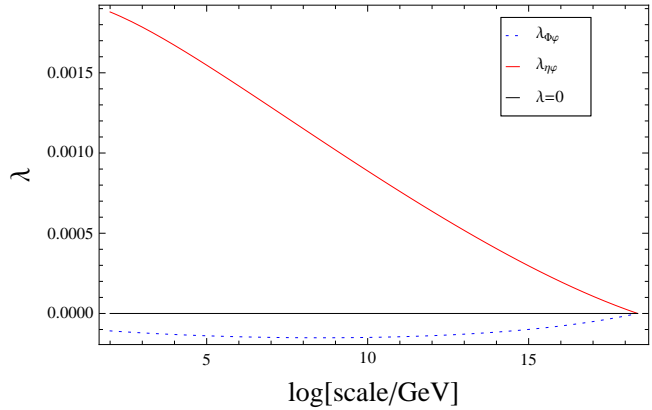}
   \caption{Running for mixings between $B-L$ Higgs and doublets. }
   \label{rge2}
\end{center}
\end{figure}
\begin{figure}[bt]
\begin{center}
\includegraphics[scale=0.7]{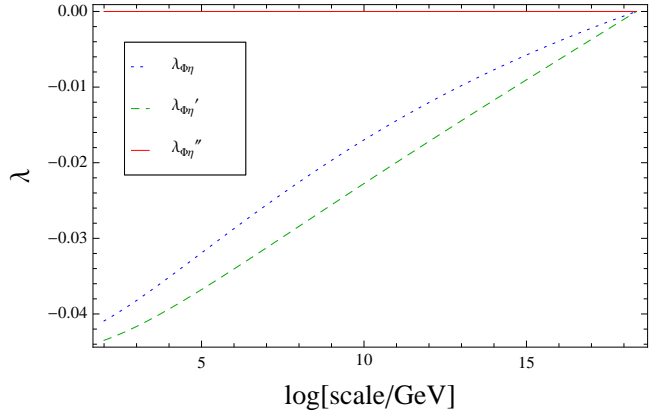}
   \caption{Running for mixings between two doublets. }
   \label{rge3}
\end{center}
\end{figure}
In Fig. \ref{rge1}, $\lambda_\varphi$ becomes negative and satisfies Coleman-Weinberg condition (see Eq. (\ref{CWcon})) 
at $v'=$10.9 TeV. At this scale, other couplings satisfy inert conditions. 
In this case, Z' mass becomes 3.7 TeV, while the experimental search for the Z' boson at LHC gives the lower bound on Z' boson mass, $m_{Z'}\geq 3$ TeV~\cite{Aad:2014cka, CMS:2013qca}. Therefore it satisfies the experimental condition. 

We investigate the LFV processes. 
In our model, we obtain $\mathrm{Br}\left(\mu\to e\gamma\right)=4.6\times 10^{-14}$, 
$\mathrm{Br}\left(\mu\to eee\right)=3.3\times 10^{-16}$ and the conversion rates~\cite{Vicente:2014wga} 
$CR(\mu-e, Ti)=1.4\times 10^{-15}$, $CR(\mu-e, Au)=7.7\times 10^{-16}$. 
The most stringent experimental upper bound of the branching ratio is
$\mathrm{Br}\left(\mu\to e\gamma\right)=4.2\times 10^{-13}$~\cite{meg3} 
and the other experimental upper bounds are $\mathrm{Br}\left(\mu\to eee\right)=2.7\times 10^{-8}$~\cite{Hayasaka:2010np}, 
$\mathrm{CR}(\mu-e, Ti)=4.3\times 10^{-12}$~\cite{Dohmen:1993mp}, $\mathrm{CR}(\mu-e, Au)=7\times 10^{-13}$~\cite{Bertl:2006up}. 
Therefore we can avoid any LFV processes.

\section{Dark matter}

$\eta_R$ is in favor of being a dark matter (DM) candidate in our model,
since we assume the coupling $\lambda_{\eta\varphi}$ that provides a dominant contribution to the $\eta$ mass 
 is small in our RGE result as can be seen the red line in Fig.~\ref{rge2}.
The nature is similar to the one in the original Ma model, {\it i.e.}, the pole point on the half mass of the CP even Higgses, 
or the range at around or greater than ${\cal O}$(500) GeV~\cite{Hambye:2009pw}.

However since all the parameters are uniquely fixed at the $B-L$ breaking scale, the physical values related to DM are also fixed as follows:
\begin{align}
(M_X)\equiv m_R&\approx m_I=312\ {\rm GeV},\quad m_\eta=314\ {\rm GeV}.
\label{eq:eta-mass}
\end{align}
Obviously our DM candidate cannot satisfy the measured relic density according to the above discussion.
Therefore we need to reanalyze our model so that our benchmark point can also satisfy the current relic density of the DM candidate,
or we just rely on another source of the DM candidate by assuming our DM candidate can be a partial component of DM.
To achieve the former case is technically difficult. Hence we just assume our DM is  a partial component and quantitatively estimate 
the relic density of our DM below. 
The dominant annihilation process is $2X\to2Z$, the second one is  $2X\to W^+W^-$, the third  one is  $2X\to2h_{\rm SM}$, and the last one is  $2X\to f\bar f$,
where $f$ represents the SM fermion such as top quark.
And each of  the cross section is numerically given by
\begin{align}
\sigma v_{\rm rel}(2X\to 2Z)\approx\frac{1.91\times 10^{-5}}{[\rm GeV]^2},\\
\sigma v_{\rm rel}(2X\to W^+W^-)\approx \frac{4.29\times 10^{-6}}{ [\rm GeV]^2},\\
\sigma v_{\rm rel}(2X\to 2h_{\rm SM})\approx\frac{4.29\times 10^{-10}}{ [\rm GeV]^2},\\
\sigma v_{\rm rel}(2X\to f\bar f)\approx\frac{1.74\times 10^{-11}}{ [\rm GeV]^2}.
\end{align}
Then our relic density is estimated as
\begin{align}
\Omega h^2_{X}\approx 10^{-5},\quad \frac{\Omega h^2_{X}}{\Omega h^2_{\rm total}}\approx \frac{10^{-5}}{0.12}\approx8.4\times 10^{-3}\ \%.
\end{align} 
Therefore our DM occupies $8.4\times 10^{-3}$ \% in the whole amount of DM.~\footnote{With more serious analysis, coannihilation processes have to be taken into account, since three fields are degenerated in Eq.(\ref{eq:eta-mass}). But its deviation from the annihilation result is at most ${\cal O}(10)$ \%. Therefore the situation does not change drastically.}

The spin independent elastic cross section with proton $\sigma_p$  is also obtained through
the SM Higgs portal and its value is 
\begin{align}
\sigma_p \approx \sigma_p(\Omega h^2_{\rm total})\times \left(\frac{\Omega h^2_{X}}{\Omega h^2_{\rm total}}\right) \approx 4.48 \times 10^{-46} \ [{\rm cm^2}] \times \left(\frac{\Omega h^2_{X}}{\Omega h^2_{\rm total}}\right). 
\end{align}
Thus it is completely safe for the direct detection experiment, since the strongest bound  is ${\cal O}(10^{-45})$~\cite{Akerib:2013tjd}.

\section{Conclusions}
We have investigated a classically conformal radiative neutrino model with gauged $B-L$ symmetry, in which we have successfully obtained the 
$B-L$ symmetry breaking through the Coleman-Weinberg mechanism. As a result, Majorana mass term is generated and
EW symmetry breaking occurs. We have also shown a benchmark point to satisfy several constraints such as 
inert conditions, Coleman-Weinberg condition, lepton flavor violations (especially  $\mu\to e\gamma$),
the current bound on the $Z'$ mass at LHC, and the neutrino oscillation experiments.


\section*{Acknowledgments}
\vspace{0.5cm}
Author thanks to Dr. Kei Yagyu for fruitful discussions.
This work was supported by the Korea Neutrino Research Center which is established by the National Research Foundation of Korea(NRF) grant funded by the Korea government(MSIP) (No. 2009-0083526).

\begin{appendix}
\section{RGE}
\label{sec:rge}
In this section, we analyze the RGEs at one-loop level.
The covariant derivative can be written as 
\begin{eqnarray}
D_\mu=\partial_\mu-ig'Q^YB_\mu-i\left(g_{mix}Q^Y+g_{B-L}Q^{B-L}\right)B'_\mu
  -ig\frac{\sigma^\alpha}{2}W^\alpha_\mu-ig_3T^\alpha G^\alpha_\mu, 
\end{eqnarray}
where $B_\mu$ and $B'_\mu$ are gauge bosons of $U(1)_Y$ and $U(1)_{B-L}$, 
and $Q^Y , Q^{B-L}$ are their charge operators.  
The RGE formulae for the gauge couplings are 
\begin{eqnarray}
(4\pi)^2\frac{dg'}{dt}=7g'^3, 
\end{eqnarray}
\begin{eqnarray}
(4\pi)^2\frac{dg}{dt}=-3g^3, 
\end{eqnarray}
\begin{eqnarray}
(4\pi)^2\frac{dg_3}{dt}=-7g_3^3, 
\end{eqnarray}
\begin{eqnarray}
(4\pi)^2\frac{dg_{B-L}}{dt}=g_{B-L}\left(12g_{B-L}^2+\frac{32}{3}g_{B-L}g_{mix}
 +7g_{mix}^2\right), 
\end{eqnarray}
\begin{eqnarray}
(4\pi)^2\frac{dg_{mix}}{dt}=12g_{B-L}^2g_{mix}
 +\frac{32}{3}g_{B-L}\left(g_{mix}^2+g'^2\right)+7g_{mix}\left(g_{mix}^2+2g'^2\right). 
\end{eqnarray}

The RGE formulae for the quartic couplings are given by
\begin{align}
(4\pi)^2\frac{\lambda_{\Phi}}{dt}&=
24\lambda_{\Phi}^2 +2 \lambda_{\Phi\eta}^2 + \lambda'^2_{\Phi\eta}  + 4 \lambda''^2_{\Phi\eta}
+2 \lambda_{\Phi\eta}  \lambda'_{\Phi\eta} + \lambda^2_{\Phi\varphi} \nn\\
&+\frac{3}{8}\left[2g^4+\left(g^2+g'^2+g^2_{mix}\right)^2\right]
-3\lambda_{\Phi}\left[  3g^2 + g'^2 + g_{mix}^2  \right] -6y^4_t + 12 \lambda_\Phi y^2_t,
\end{align}
\begin{align}
(4\pi)^2\frac{\lambda_{\eta}}{dt}&=
24\lambda_{\eta}^2 +2 \lambda_{\Phi\eta}^2 + \lambda'^2_{\Phi\eta}  + 4 \lambda''^2_{\Phi\eta}
+2 \lambda_{\Phi\eta}  \lambda'_{\Phi\eta} + \lambda^2_{\eta\varphi} 
+\frac{3}{8}\left[2g^4+\left(g^2+g'^2+g^2_{mix}\right)^2\right]\nn\\
&-3\lambda_{\eta}\left[  3g^2 + g'^2 + g_{mix}^2  \right] 
-2Tr\left[y_\eta^\dagger y_\eta y_\eta^\dagger y_\eta\right] 
+ 4 \lambda_\eta Tr\left[y_\eta^\dagger y_\eta \right],
\end{align}
\begin{align}
(4\pi)^2\frac{\lambda_{\varphi}}{dt}&=
20\lambda_{\varphi}^2 +2 ( \lambda^2_{\Phi\varphi} + \lambda^2_{\eta\varphi}  ) 
+96 g^4_{B-L} -48\lambda_\varphi g^2_{B-L}
-Tr\left[y_N^\dagger y_N y_N^\dagger y_N\right] 
+ 2 \lambda_\varphi Tr\left[y_N^\dagger y_N \right],
\end{align}
\begin{align}
(4\pi)^2\frac{\lambda_{\Phi\eta}}{dt}&=
 \lambda_{\Phi\eta}
 \left[
 4 \lambda_{\Phi\eta}
+
12\lambda_{\Phi}
+ 
12\lambda_{\eta}
+2Tr\left[y_\eta^\dagger y_\eta + y_\ell^\dagger y_\ell\right] 
-3\left(3g^2+g'^2+g^2_{mix} \right) +6y_t^2
\right]\nn\\
&
+2\lambda_{\Phi\varphi}\lambda_{\eta\varphi}
+4\lambda_{\eta}\lambda'_{\Phi\eta}
+4\lambda_{\Phi}\lambda'_{\Phi\eta}
+2\lambda'^2_{\Phi\eta}
+8\lambda''^2_{\Phi\eta}
+\frac{3}{4}\left(2g^4+\left(g^2-g'^2-g_{mix}^2\right)^2\right)\nn\\
&-4 Tr\left[y_\eta^\dagger y_\eta y_\ell^\dagger y_\ell\right],
\end{align}
\begin{align}
(4\pi)^2\frac{\lambda'_{\Phi\eta}}{dt}&=
 \lambda'_{\Phi\eta}
 \left[
 4\lambda_{\Phi}
+
 4\lambda_{\eta}
+
8 \lambda_{\Phi\eta}
+
 4\lambda'_{\Phi\eta}
+2Tr\left[y_\eta^\dagger y_\eta + y_\ell^\dagger y_\ell\right] 
+6 y_t^2 
-3\left(3g^2+g'^2+g^2_{mix} \right) 
\right]\nn\\
&
+16 \lambda''^2_{\Phi\eta}
+3g^2\left(g'^2+g_{mix}^2\right)
+ 4 Tr\left[y_\eta^\dagger y_\eta y_\ell^\dagger y_\ell\right],
\end{align}
\begin{align}
(4\pi)^2\frac{\lambda''_{\Phi\eta}}{dt}&=
4 \lambda''_{\Phi\eta}
 \left[
 \lambda_{\Phi}
+
 \lambda_{\eta}
+
2 \lambda_{\Phi\eta}
+
3 \lambda'_{\Phi\eta} 
+\frac12Tr\left[y_\eta^\dagger y_\eta + y_\ell^\dagger y_\ell\right] 
+\frac32 y_t^2 
-\frac{3}{4}\left(3g^2+g'^2+g^2_{mix} \right) 
\right],
\end{align}
\begin{align}
(4\pi)^2\frac{\lambda_{\Phi\varphi}}{dt}&=
4  \lambda_{\Phi\varphi}^2
+12 \lambda_{\Phi\varphi} \lambda_{\Phi}
+(4 \lambda_{\Phi\eta} +2  \lambda'_{\Phi\eta}) \lambda_{\eta\varphi}
+8 \lambda_{\Phi\varphi} \lambda_{\varphi} +12 g^2_{mix}g^2_{B-L}
\nn\\
&+\lambda_{\Phi\varphi}\left[6y^2_t+Tr\left[y_N^\dagger y_N \right]
-\frac{3}{2}\left(3g^2+g'^2+g^2_{mix} \right) -24 g^2_{B-L}
\right]
,
\label{RGEphivphi}
\end{align}
\begin{align}
(4\pi)^2\frac{\lambda_{\eta\varphi}}{dt}&=
4  \lambda_{\eta\varphi}^2
+12 \lambda_{\eta\varphi} \lambda_{\eta}
+(4 \lambda_{\Phi\eta} +2  \lambda'_{\Phi\eta}) \lambda_{\Phi\varphi}
+8 \lambda_{\eta\varphi} \lambda_{\varphi}
+12 g^2_{mix}g^2_{B-L}
-4 Tr\left[y_\eta^\dagger y_\eta y_N^\dagger y_N\right],
\nn\\
&+\lambda_{\eta\varphi}\left[6y^2_t+Tr\left[y_N^\dagger y_N \right]
-\frac{3}{2}\left(3g^2+g'^2+g^2_{mix} \right) -24 g^2_{B-L}
\right]. 
\end{align}

The RGE for the Yukawa couplings are given by 
\begin{align}
(4\pi)^2\frac{dy_\eta}{dt}
&=
y_\eta\left[\frac{3}{2}y_\eta^\dagger y_\eta + \frac{1}{2}y_\ell^\dagger y_\ell
 +Tr\left[y_\eta^\dagger y_\eta\right]
 -\frac34\left(g'^2+g_{mix}^2\right)-\frac94g^2-6g_{B-L}^2-3g_{B-L}g_{mix}\right],
\end{align}
\begin{align}
(4\pi)^2\frac{dy_\ell}{dt}
&=
y_\ell\left[\frac{3}{2}y_\ell^\dagger y_\ell + \frac{1}{2}y_\eta^\dagger y_\eta
 +Tr\left[y_\ell^\dagger y_\ell\right]
 -\frac{15}{4}\left(g'^2+g_{mix}^2\right)-\frac94g^2-6g_{B-L}^2-9g_{B-L}g_{mix}\right],
\end{align}
\begin{align}
(4\pi)^2\frac{dy_t}{dt}
&=
y_t\left[\frac92y_t^2-8g_3^2 -\frac94g^2
 -\frac{17}{12}\left(g'^2+g_{mix}^2\right)-\frac23g_{B-L}^2
 -\frac53g_{mix}g_{B-L}\right],
\end{align}
\begin{align}
(4\pi)^2\frac{dy_N}{dt}
&=
y_N\left[y_N^\dagger y_N+\frac{1}{2}Tr\left[y_N^\dagger y_N\right]
 -6g_{B-L}^2\right].
\end{align}

\end{appendix}

\end{document}